# Continuous transition from weakly localized regime to strong localization regime in $Nd_{0.7}La_{0.3}NiO_3$ films

Ravindra Singh Bisht[*1], Gopi Nath Daptary[2], Aveek Bid[2], and A. K. Raychaudhuri[1]

1. Department of Condensed Matter Physics and Material Sciences, S.N. Bose National Centre for Basic Sciences, Block JD, Sector III, Salt Lake, Kolkata 700106, India
2. Department of Physics, Indian Institute of Science, Bangalore 560012, India

*E-mail: ravindra.bisht@bose.res.in

## Abstract

We report an investigation of Metal-Insulator Transition (MIT) using conductivity and magnetoconductance (MC) measurements down to 0.3 K in $Nd_{0.7}La_{0.3}NiO_3$ films grown on crystalline substrates of $LaAlO_3$ (LAO), $SrTiO_3$ (STO), and $NdGaO_3$ (NGO) by pulsed laser deposition. The film grown on LAO experiences a compressive strain and shows metallic behavior with the onset of a weak resistivity upturn below 2 K which is linked to the onset of weak localization contribution. Films grown on STO and NGO show a cross-over from a Positive Temperature Coefficient (PTC) resistance regime to Negative Temperature Coefficient (NTC) resistance regime at definite temperatures. We establish that a cross-over from PTC to NTC on cooling does not necessarily constitute a MIT because the extrapolated conductivity at zero temperature ($\lim_{T \to 0} \sigma = \sigma_0$) though small (<10 S/cm) is finite, signalling the existence of a bad metallic state and absence of an activated transport. The value of $\sigma_0$ for films grown on NGO is reduced by a factor of 40 compared to that for films grown on STO. We show that a combination of certain physical factors makes substituted nickelate (that are known to exhibit first-order Mott type transition), undergo a continuous transition as seen in systems undergoing disorder/composition driven Anderson transition. The MC measurement also support the above observation and show that at low temperature there exists a positive MC that arises from the quantum interference which co-exists with a spin-related negative MC that becomes progressively stronger as the electrons approach a strongly localized state in the film grown on NGO.

Keywords: Continuous Transition, Disorder, Strain, Fermi/Non-Fermi Liquid and Bad Metal

## I. Introduction

Perovskite rare earth nickelates (RNiO$_3$) are well-researched strongly correlated systems which show a temperature driven MIT for all the rare earth elements except Lanthanum (La) [1, 2]. LaNiO$_3$ shows a metallic behavior at all temperatures, except in ultra-thin films of LaNiO$_3$ where a temperature driven MIT has been observed [3, 4]. It has been observed that by changing the size of rare earth ion (R), one can tune the metal-insulator transition temperature (T$_{MI}$), which is enhanced as the ionic radius is reduced [2, 5]. The phase diagram of the material with ions of different ionic radii at the R site also has impact on the spin order. For instance, when the ion radii is relatively larger ionic radii like Pr and Nd show Antiferromagnetic order (AFM) along with the MIT. For R ions with even lower ionic radii the AFM transition gets de-coupled from the MIT and the Neel Temperature T$_N$ < T$_{MI}$ [2].

The T$_{MI}$ in RNiO$_3$ can also be tuned by a number of physical parameters like substrate induced strain, lattice symmetry, hydrostatic pressure, carrier doping and oxygen vacancies etc. [2, 5, 6-14]. The MIT in RNiO$_3$ is accompanied by a structural transition from high temperature orthorhombic (Pbnm) to monoclinic (P$_{21/n}$) phase [2, 5, 15]. The metallic LaNiO$_3$ is rhombohedral (R3C) throughout the entire temperature range.

Substitution of La in Nd (Nd$_{1-x}$La$_x$NiO$_3$) poses an interesting set of questions. For x below a certain La concentration x$_c$ (x$_c \approx 0.35$ discussed below) the finite T$_{MI}$ is sustained as expected for a temperature driven MI transition. However, for x$_c \geq 0.35$ the transition may take the characteristics of a composition driven MI transition as seen for Anderson transition. It is expected that for x $\approx$ x$_c$ the transition would show features that can have features of both and issues like weak localization and quantum corrections to conductivity will appear to contribute along with Mott type temperature driven transition. This is also the regime where along with strain, factors like quenched disorder will play a role. The MIT in substituted Nd$_{1-x}$La$_x$NiO$_3$ also has a structural basis. The system is orthorhombic for x $\leq$ 0.4 while rhombohedral for x > 0.4. It is expected that in films with composition close to x$_c$, existence of orthorhombic distortion may lead to high resistance, if not stabilization of the insulating phase. Recently role of large quenched disorder on MIT in La substituted RNiO$_3$ has been investigated. Role of quenched disorder in films of La$_{0.5}$Eu$_{0.5}$NiO$_3$ (LENO) shows a large drop in resistivity across the MIT without affecting the T$_{MI}$. The Fermi Liquid behavior is retained even if resistivity is very large in this substituted nickelates [11].

In the RNiO$_3$ series, NdNiO$_3$ shows a clear temperature driven first-order MIT, which in the bulk sample occurs at T$_{MI}$ ≈ 200 K. T$_{MI}$ can be strained tuned by growing the films on appropriate substrates [2, 5, 9, 12]. When the film is grown on substrates like LAO which leads to compressive strain, T$_{MI}$ can be suppressed down to 100 K [2, 14]. On the other hand, lower symmetry substrates like NGO creates a tensile strain and T$_{MI}$ can vary from 150 K to 335 K depending on the substrate orientation [5, 12]. A recent study on NdNiO$_3$ films grown on NGO substrate of different orientation has shown that amongst (001), (110) and (100) orientation of NGO substrate, the highest orthorhombic distortion imposed by the NGO (100) substrate on NdNiO$_3$ films lead to enhancement of the T$_{MI}$ by a factor of nearly 140 K [12]. Growth of a film on such substrate as NGO (100) thus allows us to study effect of substrate induced orthorhombic distortion along with strain and quenched disorder close to the critical composition range of the system Nd$_{1-x}$La$_x$NiO$_3$ as has been discussed before.

The MIT in substituted Nd$_{1-x}$La$_x$NiO$_3$ as x is varied has been investigated. The temperature dependent resistivity measurements show that Nd$_{1-x}$La$_x$NiO$_3$ (x > 0.4) system is completely metallic with a temperature driven metallic to semiconductor transition for x ≤ 0.35 [16- 17]. Results on Nd$_{1-x}$La$_x$NiO$_3$ thin films suggests that the MIT looks similar to their bulk counterpart, but a change in the lattice constant and the Ni-O bond angle shifts the T$_{MI}$ to a lower temperature [17]. The present study has been principally motivated by the identification of metallic and insulating state in Nd$_{1-x}$La$_x$NiO$_3$ when x is close to x$_c$ and modifications that occur due to strain, quenched disorder (due to substitution) and orthorhombic distortion.

Earlier experiments on RNiO$_3$ system [3, 9, 16, 17] have identified the transition temperature T$_{MI}$ as a cross-over from a PTC ($\frac{d\rho}{dT} > 0$) to a NTC ($\frac{d\rho}{dT} < 0$) of resistivity. For many cases this identification of MIT may indeed be correct, but system like Nd$_{1-x}$La$_x$NiO$_3$ where the insulating state is severely suppressed due to the presence of La, the crossover from PTC to NTC does not necessarily mean MIT. The pristine definition of the insulating state is that the conduction is activated and an absence of conduction by extended state would mean $\lim_{T \to 0} \sigma \to$ 0. In case of many disordered oxides (as in many semiconductors as well) [18, 19] there are enough evidences to suggest that a crossover from PTC to NTC region may occur when $\sigma < \sigma_{Mott}$ [18-23]. However, this does not necessarily imply entry to an insulating state and this crossover temperature cannot be identified with T$_{MI}$ without any qualification. This issue assumes particular renewed significance in doped NdNiO$_3$ as in Nd$_{1-x}$La$_x$NiO$_3$, where there exists a critical concentration ($x_c$) and it becomes metallic for $x > x_c$. The question that we would like to raise and investigate is whether the transition from PTC to NTC close to $x_c$ can be a continuous transition and the crossover from a PTC to NTC at some finite temperature $T^*$ need not necessarily signify a MIT as one would see in a Mott type of first order transition.

The substitution driven continuous MIT has been studied in oxides like $LaNiO_3$, $SrRuO_3$, $LaNi_{1-x}Co_xO_3$, $NaOsO_3$, and $Na_xWO_3$ [24-29]. It has been seen that the role of disorder, reduced dimensionality, and renormalization of electron-electron interactions at lower temperature can give rise to an upturn in the resistivity which may be considered as quantum correction to the conductivity [23-27]. In such system, the tunneling conductance (that measures the Density of states at the Fermi level) can become temperature dependent due to presence of correlations. This has been predicted theoretically and also observed in oxides as well as doped semiconductors [30]. It has been seen that quantum confined $LaNiO_3$ shows such a transition from first-order MIT to Anderson transition [26]. Theoretical investigations have also raised the possibility that disorder in a Mott insulator can tune the MIT to a continuous transition [31, 32]. The disorder induced continuous transition has been proposed in the context of a half-filled Anderson Hubbard model at a finite temperature where the quenched disorder, as well as auxiliary field fluctuations, is responsible for Non-Fermi Liquid scaling [31].

In this paper, the issue of continuous transition for the substituted $Nd_{1-x}La_xNiO_3$ (x = 0.3) system has been investigated in the context of effects of physical factors like strain, lattice symmetry, and quenched disorder that can tune the nature of the transition. The particular composition being used here $Nd_{1-x}La_xNiO_3$ (x = 0.3) is at the verge of a composition driven insulator-metal transition. In the bulk, the sample with x = 0.3 shows a metallic behavior at a higher temperature and an upturn in resistivity at lower T [16]. The resistivity data down to 4.2 K was interpreted using variable range hopping, indicating an insulating state. Thin films (8 Unit Cell) of $Nd_{1-x}La_xNiO_3$ (x = 0.25) grown on LAO shows a metallic resistivity down to 75 K and an upturn in resistivity below 75 K [17]. If upturn in resistivity is taken as $T_{MI}$, the phase diagram of films of $Nd_{1-x}La_xNiO_3$ suggests that $x^* \sim 0.35$[17]. In both the studies upturn in resistivity was identified as $T_{MI}$. From the $x - T_{MI}$ phase diagram, it appears that the composition x = 0.3 though close to the composition driven MIT, is on the insulating side.

The investigation has been done on films of $Nd_{1-x}La_xNiO_3$ (x=0.3) grown on substrates LAO, STO and NGO. The crystal structures of substrates used LAO (100), STO (100), and NGO (100) are rhombohedral, cubic, and orthorhombic respectively. The main motivation for choosing different substrate are: (1). To vary the amount of built-in strain that range from being compressive to tensile, (2). The nature of in-plane strain that is isotropic for films grown on LAO and STO due to their cubic nature while anisotropic for film grown on NGO, and (3). The orthorhombic distortion imposed by the NGO (100) which can push the transition temperature to a higher value and it has been already reported for $NdNiO_3$ [12]. The question we have asked is that even if we tune the transition to somewhat higher temperature (by inducing orthorhombic distortion using NGO (100)), does the nature of transition retain its first order character or it shows a metallic state albeit with very low $\sigma$ $(T = 0)$ but finite.

The films have a thickness $\sim 80$ nm. At this thickness the films are not coherently strained, but even at such thickness the transition does not approach the bulk value. The substrate mismatch leads to enough residual strain even if the films are partially strain relaxed [2, 8, 11, 12, 33, 34]. From measurements of conductivity down to 0.3 K, it has been shown that in $Nd_{0.7}La_{0.3}NiO_3$ film (here after referred as NLNO), there is a definite cross over temperature $(T^*)$ that depends on the substrate on which the film is grown. In the low temperature region $(T \ll T^*)$, the NLNO film is found to be a "bad metal" with extremely low but finite value of the conductivity so that $\lim_{T \to 0} \sigma(T) \neq 0$. This raises the possibility of a continuous MIT in the $Nd_{0.7}La_{0.3}NiO_3$ system in contrast to a first order MIT in $NdNiO_3$ (NNO).

We also show that in such a system at low temperature $(T < 2$ K), a small but finite magnetoconductance (MC) develops. The MC ( $\Delta\sigma(H) \equiv \sigma(H) - \sigma(H = 0)$) becomes non-trivial and shows signature for formation of local moments that co-exists with contribution from electron localization. Scanning Tunneling Spectroscopy (STS) data (taken down to $T^*$) show that the films of NLNO grown on different substrates, have a metallic density of state (DOS) at the Fermi level, $N(E_F)$. However, there is a distinct suppression of the $N(E_F)$ in the film grown on NGO that shows bad metallic behavior with a very small $\sigma(T = 0)$, as the temperature is lowered towards the cross-over temperature $T^*$.

## 2. Experimental

The films were made by Pulsed Laser Deposition (PLD) using well characterized pellet of NLNO (made by sol-gel process) as a target using KrF ($\lambda$ = 248 nm) laser of pulse duration 10 ns. The films were grown on rhombohedral LaAlO$_3$ (100), cubic SrTiO$_3$ (100), and orthorhombic NdGaO$_3$ (100) single crystal substrates using a laser Fluence 2.5 J/cm$^2$ at a repetition rate of 5 Hz. The PLD chamber was evacuated down to $10^{-6}$ mbar and purged with high purity oxygen before the deposition. The substrate temperature and oxygen pressure during deposition was 675$^0$C and 0.1 mbar respectively. Post-deposition, the films were annealed for 30 minutes at 675$^0$C in 1 atmospheric oxygen pressure and slowly cooled down to room temperature. The films were structurally characterized using X-ray Diffraction (XRD) $\theta - 2\theta$ scan.

The transport measurements were done with 4 co-linear probes Cr/Au pads on the sample. The resistivity was calculated using the relation $\rho = R\left(\frac{A}{L}\right)$, where R is the measured resistance, A is the cross sectional area and L is the length between two consecutive contact pads. Four Cr/Au contact pads of size $\sim$ 1.2 mm $\times$ 2.5 mm were made on the films by thermal evaporation. The sample was mounted on a chip carrier followed by wire bonding. The conductivity $\sigma(T)$ as a function of T as well as relative MC $\frac{\Delta\sigma}{\sigma}(H) \equiv \frac{\sigma(H)-\sigma(H=0)}{\sigma(H=0)}$, were measured in a He-3 cryostat in the temperature range 0.3 K – 30 K at H $\leq$ 8 T. For measurement in zero fields in the range 3 K -300 K, a cryogen free pulsed tube cryocooler was used. The Scanning Tunneling Spectroscopy (STS) measurements on the films were performed using a UHV Scanning Tunneling Microscope (STM). The base pressure of the STM was $10^{-10}$ mbar. Using a very small ac modulation over a fixed bias voltage local tunneling conductance $g = \frac{dI}{dV}$ was measured in a bias range |V| $\leq$ 0.2 V.

## 3. Results and Analysis

### 3.1. Thin film characterization:

The XRD data ($\theta - 2\theta$ scan) are shown in Fig. 1, where the intensity is plotted in log scale. The XRD data show that the films are free from minority/impurity phases. The in-plane misfit strain has been calculated according to $\epsilon_{inplane} = \frac{a_{sub} - a_{bulk}}{a_{bulk}}$. $a_{sub}$ and $a_{bulk}$ are the pseudo cubic lattice parameter of substrate and NLNO bulk respectively. The out of plane strain was calculated according to $\epsilon_{out\ of\ plane} = \frac{2\nu}{(\nu-1)}*\epsilon_{in-plane}$, where $\nu$ is the Poisson ratio. For NdNiO$_3$ $\nu = 0.25$ [35] and the out of plane strains are given as $\epsilon_{out\ of\ plane} = -0.66\epsilon_{in-plane}$. Data on the samples used are summarized in Table 1.

The XRD data show that NLNO/LAO and NLNO/STO films are strongly textured in the (110) direction with similar full width at half maxima (FWHM), while NLNO/NGO film is textured along (100) direction with relatively low value of FWHM. (The miller indices of peaks from which FWHM are calculated are given in Table 1). The different growth directions and FWHM scan be attributed to the fact that films NLNO/STO and NLNO/LAO have a cubic symmetry while the film NLNO/NGO has an orthorhombic symmetry which arises from the different crystallographic nature of the substrates. In Table 1 the relevant structural parameters like lattice parameters (a, b, c), pseudocubic lattice parameters ($a_{pc}$), crystal structures, strains, growth directions, FWHM are summarized. The in-plane strain is compressive for the film NLNO/LAO while it is tensile for the films NLNO/STO and NLNO/NGO. The in-plane strain is expected to be isotropic for the films grown on rhombohedral LAO (a = b ≠ c) and cubic STO (a = b = c) substrates. However, due to orthorhombic distortions (a ≠ b ≠ c) the in-plane strain in the NLNO/NGO film is expected to be anisotropic. The XRD establish that the substrate symmetry plays an important role in the film growth and nature of strain experienced.

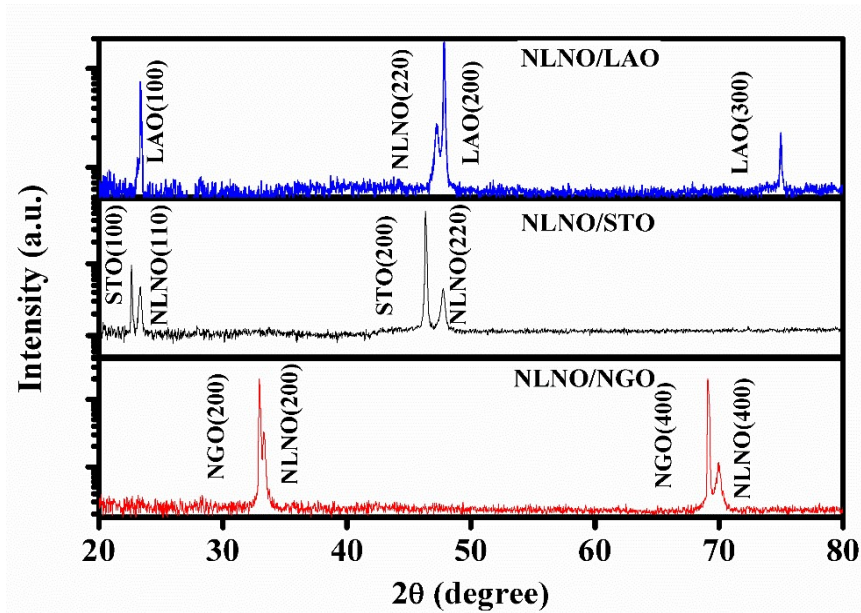

**Figure 1.** XRD $(\theta - 2\theta)$ scans of NLNO films grown on LAO, STO, and NGO single crystal substrates. The intensities are plotted in log scales.

Due to relatively larger thickness, films are not epitaxial (coherently strained) as it happens in very thin films (few unit cells). Mismatch between substrate and film lattice constants lead to interfacial strains [34]. The relaxation of interfacial strains lead to additional disorder, which adds to the quenched disorder arising from a random substitution of Nd ($1.163A°$) by La ($1.216A°$) [36]. These two quenched disorder would contribute to the electronic transport in this system.

Table 1 Structural parameters of NLNO bulk, substrates, strain and FWHM as obtained from XRD curves. The subscript in FWHM shows the miller indices for the peaks for which FWHM was calculated.

| Sample | Crystal Structure | a (Å) | b (Å) | c (Å) | a$_{pc}$(Å) | FWHM$_{(Miller\ Indices)}$ | $\epsilon_{in-plane}$ | $\epsilon_{out\ of\ plane}$ |
|---|---|---|---|---|---|---|---|---|
| NLNO(bulk) [16] | Orthorhombic | 5.403 | 5.377 | 7.644 | 3.82 | - | - | - |
| LAO | Rhombohedral | 3.79 | 3.79 | 13.11 | 3.79 | 0.27$_{(220)}$ | -0.79 | 0.52 |
| STO | Cubic | 3.905 | 3.905 | 3.905 | 3.905 | 0.30$_{(220)}$ | 2.22 | -1.46 |
| NGO | Orthorhombic | 5.43 | 5.50 | 7.71 | 3.86 | 0.18$_{(200)}$ | 1.04 | -0.68 |

### 3.2. Resistivity as a function of temperature

The resistivity ($\rho$) data for all the films are shown in a linear temperature scale in Fig. 2 in the temperature range 0.3 K – 300 K. Data have been taken in heating cycle. At room temperature, $\rho$ is the lowest in NLNO/LAO with compressive in-plane strain and highest for the film NLNO/STO which has highest built in in–plane tensile strain (Table 1). The derivative $\frac{d\rho}{dT}$ has been used to mark the crossover temperature T$^*$ from PTC to NTC behavior. The derivative $\frac{d\rho}{dT} = 0$ defines T$^*$. A clear crossover with well-marked T$^*$ has been seen for the films with tensile strains (grown on NGO and STO, marked by arrows) while for the film with compressive strain grown on LAO, no clear T$^*$ is observed. $\rho$ for NLNO/NGO at 300 K is substantially lower than that of NLNO/STO. However, due to rather large T$^* \approx$ 150 K , at lower temperature $\rho$ of the film rises rapidly on cooling and for T $<$ 50 K it becomes larger than that of the film NLNO/STO.

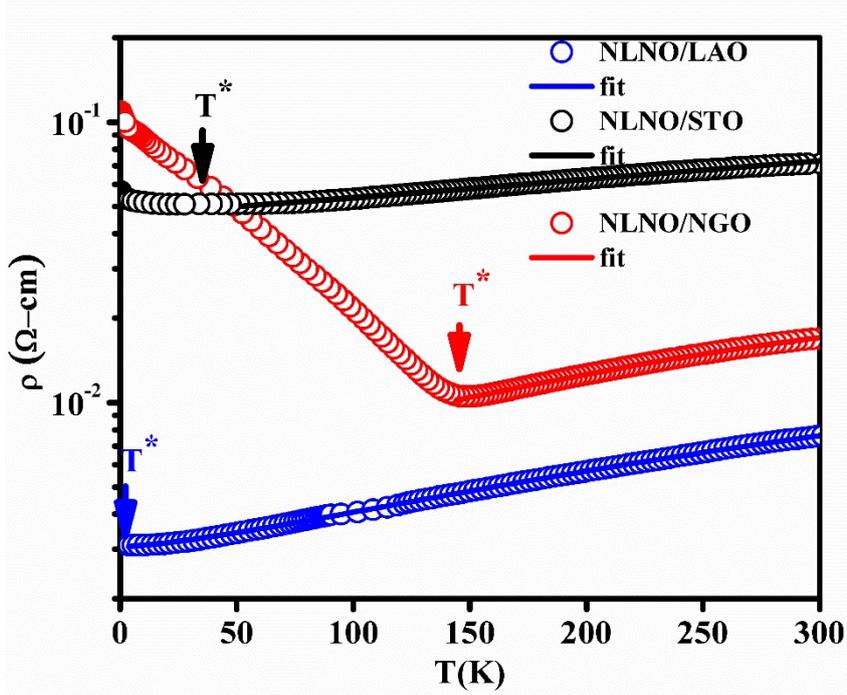

**Figure 2.** $\rho$ as a function of temperature T for NLNO films grown on LAO, STO, and NGO single crystal substrates. The fits to Eqn. 3 are shown as solid lines.

### 3.2.1 Low-temperature conductivity data:

Conductivities ($\sigma(T)$) for the three samples for the entire temperature range (0.3 K - 300 K) are shown in Fig.3. In the films with tensile strain, $\sigma(T)$ show positive temperature coefficients of conductivity ($\frac{d\sigma}{dT} > 0$) below 100 K, as has been seen at low temperature limits in transition metal oxides close to the composition driven MIT [19, 20, 26].To establish that the conductivity though low, reaches a finite value as T → 0 ($\lim_{T \to 0} \sigma(T) \neq 0$), we have plotted the derivative $W \equiv \frac{dln\sigma}{dlnT}$ vs T in the inset of Fig. 3 for T ≤ 10 K [18, 37]. For an insulator, $\lim_{T \to 0} \sigma(T) = 0$ due to activated nature of the conductivity, and W diverge as T → 0. On the contrary, when $\sigma(T)$ approaches a finite value however small it may be the $\lim_{T \to 0} \sigma(T) \neq 0$, W → 0 as T → 0 . Inset of Fig.3 shows that in all the 3 films indeed W → 0 as T → 0 signifying a finite zero temperature conductivity. This establishes that at low temperatures in all the three films $\lim_{T \to 0} \sigma(T) \neq 0$.

For NLNO/LAO $\lim_{T \to 0} \sigma(T) \sim$ 300 S/cm. In films with tensile strains, $\sigma(T)$ though finite are very low. In particular, in the film NLNO/NGO, the $\lim_{T \to 0} \sigma(T)$ <10 S/cm.

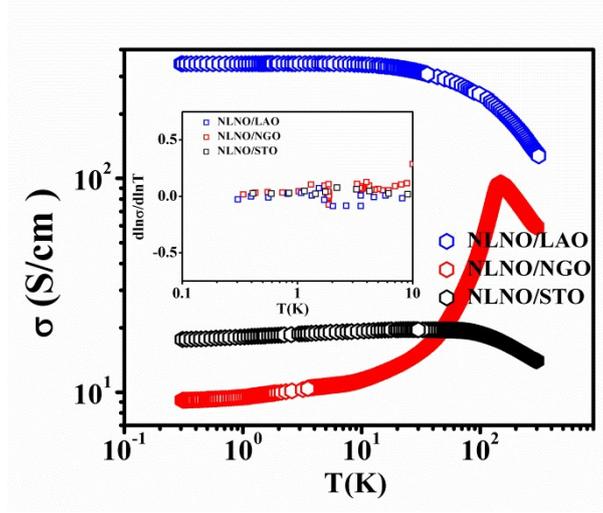

**Figure 3.** Conductivity (σ) as a function of temperature T in the three films of NLNO grown on LAO, STO, and NGO single crystal substrates. The inset shows the derivative $W \equiv \frac{d\ln\sigma}{d\ln T}$ as a function of temperature for T < 10 K.

The data thus support the hypothesis that a combination of physical factors like the strain, lattice symmetry, as well as disorder can make the MI transition continuously tunable close to a critical composition of La substitution and the first order transition seen in NNO is not observable. In general, in a disordered system, it has been shown that strain, hydrostatic pressure, as well as disorder, can tune in a continuous MIT as has been shown in Si: P system close to the critical P concentration [18, 37-40].This is also well documented for other disordered oxides [18-26]. In La substituted NLNO, approach to the MI transition thus assumes the nature of a disorder and composition driven Anderson Transition.

The approach to MIT in the low temperature regime in an Anderson transition generally follows power law in the metallic side of the transition [39]. The $\sigma(T)$ vs T data at low T (T ≤ 2 K) are fitted with a power law in linear scale (Fig. 4(a)) to derive the limiting conductivity:

$$\sigma(T) = \sigma_0 + \text{AT}^n \qquad (1)$$

(The low temperature limiting power law behavior starts in the three films at different temperatures. For NLNO/LAO it starts below 2 K. For NLNO/STO it starts below 20 K and in NLNO/NGO at somewhat higher temperatures. However, to bring them on the same basis, the power laws have been fitted at T ≤ 2 K for all the films.)The exponent $n$ evolves as $\sigma_0$ decreases on approach to the MIT. The data is very similar to that observed previously on a number of disordered oxides undergoing Anderson type MIT [18-20, 23, 25].

Figure 4(b) shows the evolution of $\sigma_0$ as a function of n, which is a parameter that measures closeness to MIT. Of particular interest is the value of $n \approx 1$ seen in the film on NGO substrate, that has the lowest value of $\sigma_0$. In oxides [19] with similarities to NNO, it has been seen that close to the MIT, the exponent $n \rightarrow 1$. Often a scale of Mott minimum metallic conductivity ($\sigma_{Mott}$) is used as a scale of conductivity that signals transition to a bad metallic state. For most oxides like NNO where the charge density is in the vicinity of $10^{22} cm^{-3}$, $\sigma_{Mott} \sim (5-10) \times 10^2 \ S/cm$. Thus for the film NLNO/LAO, $\sigma_0 \sim \sigma_{Mott}$ and for the other two films $\sigma_0 \ll \sigma_{Mott}$. The value of $\sigma_0$ is also a small fraction of the parameter $\sigma_{MIR}(\sigma_0 \ll \sigma_{MIR} \equiv \rho_{MIR}^{-1})$. This makes the metallic state below T* a bad metallic state. We can also make a comparison with $\sigma_{MIR}(\sigma_{MIR} \equiv \frac{1}{\rho_{MIR}} \approx 3000 - 5000 \ S/cm)$.

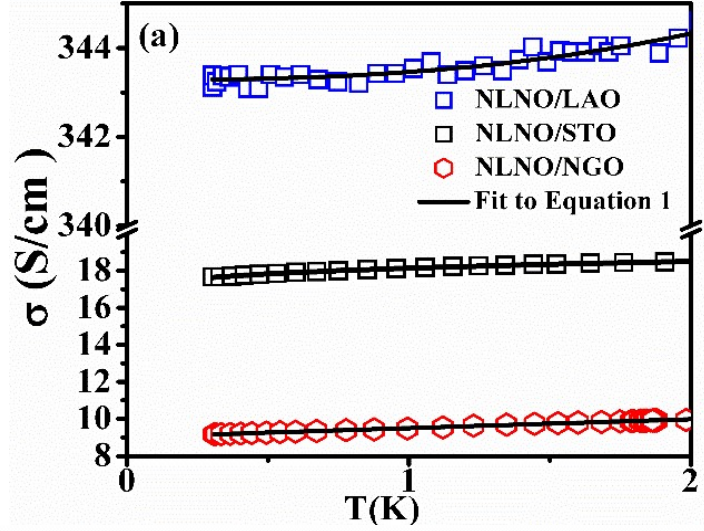

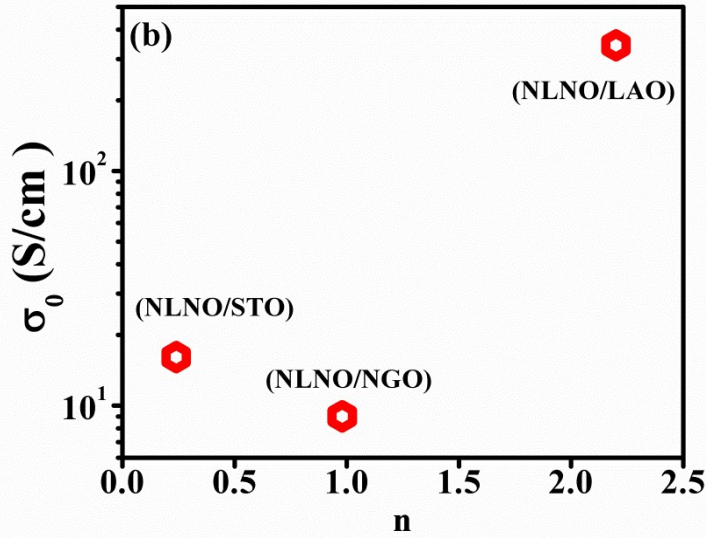

**Figure 4.** (a) The experimental data and fit to Eqn. 1 for films grown on all three substrates for the range 0.3 K ≤ T ≤ 2 K and (b) $\sigma_0$ as a function of exponent $n$ for NLNO films grown on LAO, STO, and NGO single crystal substrates.

Table 2 Parameter values for fit of conductivity data to Eqn. 1 for range 0.3 K ≤ T ≤ 2 K

| Sample | n | $\sigma_0$ (S/cm) | A (S/cmK$^{-n}$) |
|--------|---|-------------------|-------------------|
| NLNO/LAO | 2.2 | 343.0 | 0.21 |
| NLNO/STO | 0.24 | 16.1 | 2.0 |
| NLNO/NGO | 1.02 | 9.01 | 0.49 |

In NLNO/LAO which has the highest conductivity, the behavior is metallic in all regions (with PTC) and the conductivity is essentially flat below 30 K with a small $\frac{d\sigma}{dT}$ as expected from corrections to conductivity arising from weak localization[18,39]. In this regime, the correction to conductivity due to disorder is small and is given by [39]:

$$\sigma(T) = \sigma_0 + kT^{p/2} \qquad (2)$$

Where the factor $k \equiv \frac{2e^2}{\xi h \pi^2}$, $e$ is the electronic charge and $\xi$ (de-phasing length scale) is related to the Thouless length $L_{TH} = \xi T^{-p/2}$. From weak localization theory, when the interaction effect is weak $p \approx 4$ [39]. Experimentally observed $n \approx 2.2$ (Table 2) is very close to the expected value of $p/2$. The accuracy of the estimation of $n$ is limited due to very weak temperature dependence. In NLNO/STO, the exponent $n = 0.24$ is somewhat smaller than what is expected in this regime when electron-electron interaction is associated with weak localization. The value of $n$ is ½ in such cases. In this case, the de- phasing length is such that it is dominated by thermal effects. As the transition is approached, the exponent in many oxides approaches a value of $n = 1/3$, which has been explained as an exponent in the metallic side close to the critical region of MIT [18]. It may happens that the $L_{TH} >>$ the thickness of the film. In that case, the dominant length scale will be determined by the thickness, being smaller than $L_{TH}$. This may also modify the exponent. There is no theory available that explains the evolution of the exponent in such confined sample size.

### 3.2.2 High-temperature resistivity data (T > T*):

The resistivity $\rho(T)$ data in the temperature range $T > T^*$ for all the three films are fitted with a power law of T to check whether there is presence of Non-Fermi Liquid (NFL) behavior in these systems. Past studies on NNO films with compressive strain have shown the existence of NFL behavior [9, 11]. To investigate this, the data are fitted to the equation below:

$$\rho(T) = \frac{\rho^*(T)\rho_{sat}}{(\rho_{sat} + \rho^*(T))} \qquad (3)$$

Where $\rho^*(T) = \rho_0 + BT^m$. $\rho_{sat}$ is a parallel high resistance when resistivity approaches the Mott-Ioffe-Regel limit [9].The value of exponent $m$ is 2 for a Fermi Liquid (FL) and $\frac{5}{3}$ for a NFL [9]. $B$ defines the strength of electron-electron scattering. The fit to the data are shown in Fig. 2 as solid lines. The results are tabulated in Table 3.

Table 3 Parameter values for fit of the high temperature resistivity data to Eqn. 3.

| Sample | m | $\rho_{sat}$ (Ohm.cm) | $\rho_0$ (Ohm.cm) |
|---|---|---|---|
| NLNO/LAO | 1.43 | 0.060 | 0.003 |
| NLNO/STO | 2.14 | 0.086 | 0.056 |
| NLNO/NGO | 2.01 | 0.030 | 0.010 |

It can be seen that the films with tensile strain have $m \approx 2$. However, for the film NLNO/LAO, $m \approx 1.43$, which is close to $m = \frac{5}{3}$. The observed behavior in NLNO/LAO is in conformity with the observations and proposed strain-driven phase diagram for NNO [9].The observed $\rho_{sat}$ in the NLNO/STO and NLNO/LAO films are similar in order, while in NLNO/NGO, it is somewhat smaller although of the same order. $\rho_{sat}$ is proportional to the inter-band spacing, which arises due to the formation of conducting channel because of inter-band scattering [9].Typical values for Mott-Ioffe-Regel Limit ($\rho_{MIR}$) for nickelates is 0.2-0.3 m$\Omega$-cm [9, 11]. All the films here show $\rho_{sat} > \rho_{MIR}$ suggesting bad metallic nature of films. The value of $\rho_0$ seems to depend on the built-in strain. It is largest in the most strained film NLNO/STO. For the films NLNO/STO and NLNO/LAO, the observed $\rho_0$ from the high-temperature fit is very close to $\frac{1}{\sigma_0}$ obtained from the conductivity data below 2 K (see Table 2). Interestingly, due to large temperature variation of $\sigma$ in NLNO/NGO, $\rho_0 \ll \frac{1}{\sigma_0}$.

### 3.3. Tunneling conductance for$T > T^*$: How good is the metallic state?

All the three films show metallic behavior at high temperature ($T > T^*$), but they have different resistivity. The film NLNO/NGO shows a very low conducting (highly resistive) state below $T^*$. We ask the question whether we observe an anomaly in the DOS at Fermi level $E_F(N(E_F))$. We obtained a measure of $N(E_F)$ from the tunneling conductance data $g(V)$ measured as a function of bias $V$ applied to film as the tip was grounded with respect to film. The $g - V$ data for the films at room temperature and 120 K are shown in Fig. 5. For all the films, the tunneling conductance $g(V)$ has a parabolic dependence on bias $V$, as expected for tunneling of electrons between two metal electrodes (tip and the film) through a symmetric trapezoidal barrier. Any asymmetry in the trapezoidal barrier would lead to an offset ($V'$) from parabolic behavior at low bias [41].Due to asymmetric nature of the $g - V$ curve, the tunneling data are fitted to the parabolic equation that incorporates $V'$ [41]:

$$g(V) = \alpha + 3\gamma(V - V')^2 \qquad (4)$$

Where $\alpha$ is a measure of DOS, $\gamma$ is related to mean barrier height and $V'$ is the offset from parabolic behavior due to asymmetric trapezoidal barrier. The quadratic term contains a term $V'$ that accounts for any small shift of the minimum of $g(V)$ from the origin. The fit to the data are shown as solid lines and the parameters $\alpha, \gamma, V'$ are obtained. The values of $V'$ as obtained from the fit (shown in the inset of Fig. 6) are small and negative.

Deviation from parabolic tunneling equation (Eqn. 4) can occur if applied bias is comparable to the work function leading to departure from trapezoidal barrier and also if the bands from which electrons tunnel are narrow (like- d bands) so that assumption of a flat DOS cannot be made. In oxides like NNO, the d band (hybridized with Ni$^{3+}$ orbitals) has a width of ~1 eV. The tunneling conductance has been measured with a bias V < 0.2eV. At this small bias deviation from a parabolic tunneling equation is not expected. Since $\alpha$ is a measure of $N(E_F)$, $\frac{\alpha(T)}{\alpha_{293}}$ as a function of T would measure the temperature variation of $N(E_F)$. Data for the three films are shown in Fig. 6.

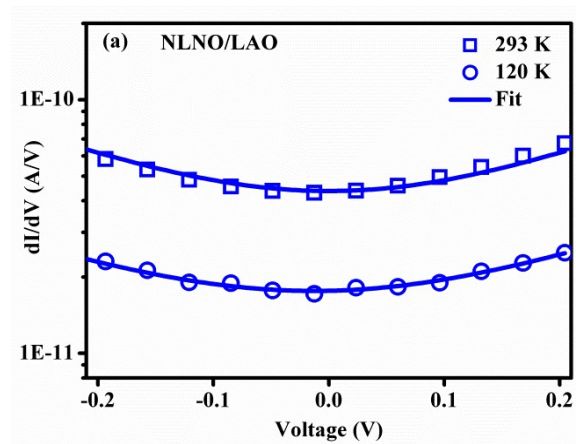

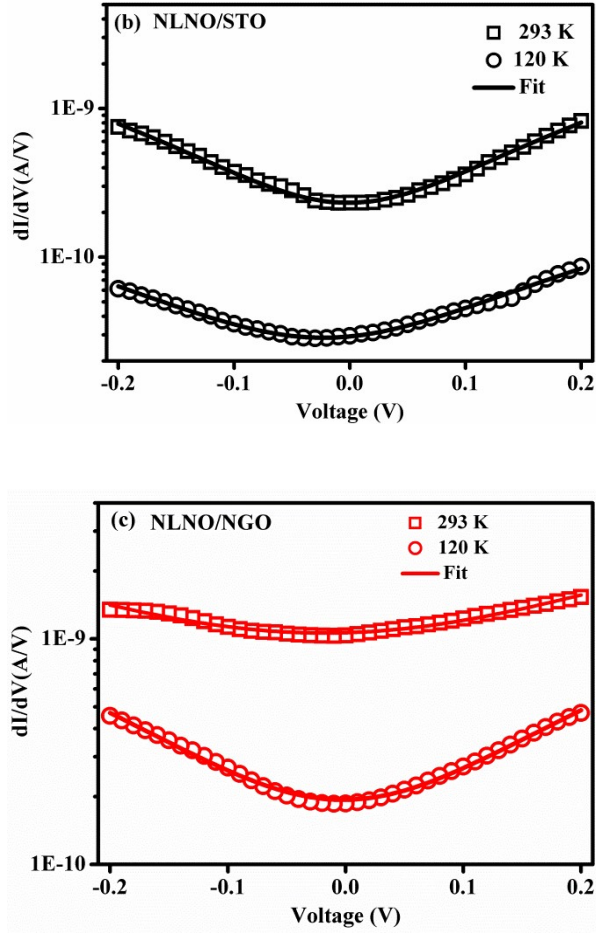

**Figure 5.** $g = \frac{dI}{dV}$ as a function of bias voltage $V$ at 120 K and 293 K for the three NLNO films. The lines through the data are fit to Eqn. 4.

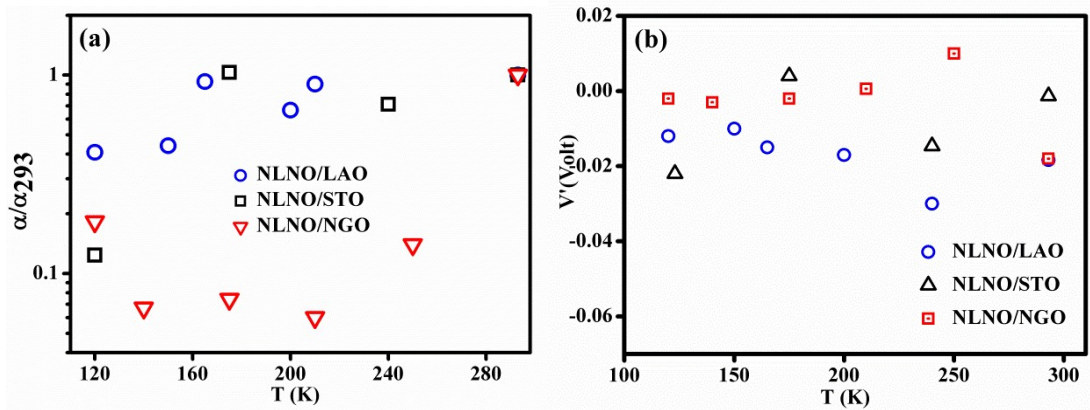

**Figure 6.** **(a)** $\frac{\alpha(T)}{\alpha_{293}}$ as a function of temperature for the three NLNO films and **(b)**The variation of V´ as a function of T in all the films. The magnitude of the term V´, however is < 20 meV.

For the NLNO/LAO, $N(E_F)$ is more or less temperature independent as it should be in a metal, barring a step decrease at around 150 K. In case of NLNO/STO, which has a distinct T*, the $N(E_F)$ starts to show a drop below 150 K as well. Interestingly, the temperature range is close to the temperature T*. In NLNO/NGO, which lies closest to the MI boundary, there is a large drop in $N(E_F)$ on cooling even in a state where the resistivity shows PTC. In a strongly correlated metal, it has been shown that due to electron-electron correlation the DOS can be depressed close to the Fermi energy $E_F$ and it also has a temperature dependence where the depression close to $E_F$ enhances on cooling [30]. The tunneling conductance data thus establish that the NLNO/NGO differs strongly from a conventional metal due to strong correlation.

### 3.4. Magnetoconductance at low temperature (T < 2 K)

A potent tool to study MIT and electron localization is MC. Two factors with opposite signs can contribute to the MC close to MIT. The quantum interference (QI) arising due to backscattering in a disordered metal leads to weak localization. QI is suppressed in an applied magnetic field due to lack of time reversal symmetry. This leads to a positive MC [42]. (This we refer as WL contribution). The other contribution arises from localized spins. Near the MIT, localization of electrons leads to formation of localized spins at energies close to the Fermi level [39, 43, 44]. The localized spins can lead to weak magnetism and can also participate in electron scattering, giving rise to negative MC. Since the contribution comes from the exchange correlation of spins at different site, we refer to this contribution as EC.

The films studied being close to the transition region (at least 2 of them), the MC is expected to have comparable contributions from both the effects. These two contributions will add-up and would give rise to a non-trivial dependence of MC on applied magnetic field ($H$). Since direct measurement of weak magnetic moments through magnetic measurements (like magnetization) in such films become difficult, the MC measurements give us an alternate way to gain information of the existence of spins in such films.

The MC data are shown in Fig. 7 at 0.3 K and 1.8 K in all the three films. The magnitudes of the MCs are small and they decrease rapidly above 2 K. At 0.3 K at a field of 8 T (data not shown), the maximum MCs observed are 1.1 %, 3.9 % and 8.0 % in the films NLNO/LAO,

NLNO/STO and NLNO/NGO respectively. In all the films, at T = 5 K, $\frac{\Delta\sigma}{\sigma} \leq 0.1\%$. For this reason, the data analysis is restricted in the range of T < 2 K.

In general, the contribution of WL to MC has a characteristic dependence $\frac{\Delta\sigma}{\sigma} \sim \sqrt{H}$ at low fields [37, 40]. The EC contribution to MC is given by the phenomenological relation $\frac{\Delta\sigma}{\sigma}(H) = -A * \frac{H^2}{H^2 + (H_e^2)}$, where $H_e$ is field required for spin alignment [41-42]. For $H \gg H_e$, $\frac{\Delta\sigma}{\sigma}(H)$ saturates. At low field $(H \ll H_e)$, $\frac{\Delta\sigma}{\sigma}(H) \sim H^2$. Assuming that the MC contributions are additive, the MC is fitted to the relation below:

$$\frac{\Delta\sigma}{\sigma}(H) = -A(T) * \frac{H^2}{H^2 + (H_e^2)} + B(T)\sqrt{H} \qquad (5)$$

Where, $A(T)$ and $B(T)$ are temperature dependent but field independent constants.

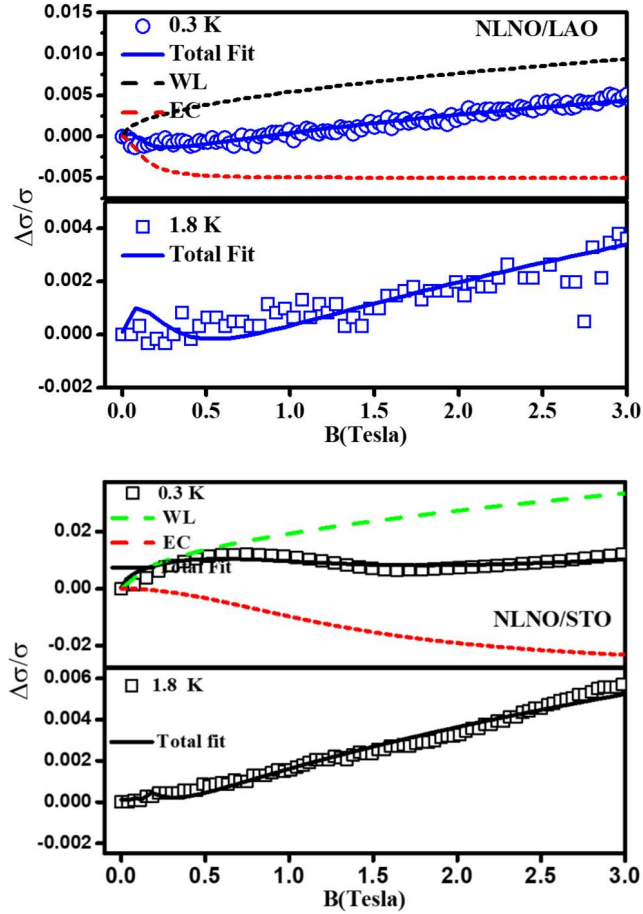

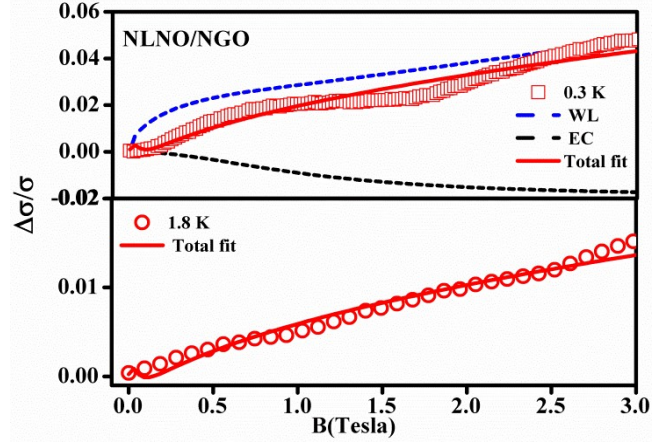

**Figure 7.** Magnetoconductance data for all the films at 0.3 K, and 1.8 K. The solid lines through the data are the fit to Eqn 5. Individual contributions are shown by dotted lines.

In Fig. 7 the two contributions are shown separately. The coefficients $A$ and $B$ decrease on heating (Table 4) leading to weakening of MC on heating. In case of WL, enhanced temperature leads to enhanced thermal de-phasing and the field induced suppression of WL becomes less effective. This also happens with the negative MC contribution because the ground state spin population decreases at higher temperature.

The values of MCs have a small dependence on the direction of field scans (field- up or field-down scans) in films NLNO/ STO and NLNO/NGO. It is likely that the spins get aligned by the random anisotropy field leading, to hysteresis in MC for the field scan-up and scan-down. In order to keep the state of the film unaffected by history of applied field, the data shown are for field scan-up direction only.

The values of $H_e$ is ~ 0.4 T, 0.05 T and 0.17 T for NLNO/LAO, NLNO/NGO and NLNO/STO respectively. In all the three films, the values of $H_e$ are temperature independent which strongly suggest that $H_e$ may arise from local anisotropy like a random field. The values for the coefficient $A$ for the films on NLNO/LAO and NLNO/STO are low. However, for the film NLNO/NGO which is on the verge of becoming insulating, the coefficient $A$ is much larger. Stronger localization of electrons in NLNO/NGO leads to enhanced effects from localized spins.

Table 4 Fitting parameter $A$ and $B$ obtained from fitting data to Eqn. 5 for the films for field scan-up H (0-3T).

| T(K) | (NLNO/LAO) | | (NLNO/STO) | | (NLNO/NGO) | |
|------|------|------|------|------|------|------|
| | **A** | **B** | **A** | **B** | **A** | **B** |
| **0.3** | 0.005 | 0.005 | 0.0028 | 0.015 | 0.012 | 0.033 |
| **1.8** | 0.0047 | 0.004 | 0.0015 | 0.005 | 0.004 | 0.010 |

## 4. Discussions

The main results obtained in this investigation establish that the resistivity upturns in films NLNO/STO and NLNO/NGO below temperature T* do not mean entry into an insulating state, rather there is an entry into a metallic phase with NTC but a finite $\sigma_0 \ll \sigma_{Mott}$. The results also establish that even with same value of $x \approx 0.3$ in NLNO the limiting resistivity at T → 0, can be tuned by a factor of nearly 40 by changing strain, lattice symmetry and the orientation of the films grown. In particular, the presence of orthorhombic distortion (as in the film NLNO/NGO) appears to have a strong effect that it almost drives the film to the critical region of MIT. The observed MIT has the features of an Anderson transition (continuous transition) in contrast to first order Mott type transition seen in NNO. The La substitution in NNO close to the critical composition thus changes the nature of the transition. The $T_{MI}$ in this case turns into a cross-over temperature that marks the change from a PTC region to a NTC region.

Theoretical investigations on half-filled Anderson Hubbard model have shown that the resistivity scaling strongly depends on the strength of interaction and disorder [30]. Also the quantum corrections to conductivity at lower temperature has been seen in disordered LaNiO₃ system and modeled as electron-electron interaction and 3-D weak localization [26].

In the three films, the effect of strain, as well as crystal lattice symmetry, play a dominant role in deciding the nature of the low T transport. The higher value of $\rho_{sat}$ in NLNO/STO compared to that of $\rho_{sat}$ in NLNO/NGO can be understood in terms of strain induced $e_g$ orbital splitting [9]. Table 1 shows the magnitude of strain is highest for NLNO/STO. Higher tensile strain in the film NLNO/STO leads to higher $\rho_{sat}$. If the magnitude of strain is solely responsible for T* then one would expect higher T* for NLNO/STO. However, the crossover temperature is less than that of the NLNO/NGO. Due to higher T* in NLNO/NGO films, its resistance starts to rise in this film much faster than that of NLNO/STO films when cooled to lower temperatures and eventually in NLNO/NGO the conductivity $\sigma_0$ becomes lower than that of NLNO/STO by a factor of nearly 1.7. The existence of higher T* in NLNO/NGO likely happens due to substrate induced lattice orthorhombic distortion which is consistent with the earlier results on NdNiO₃ [12]. It is clear that the substrate induced orthorhombic distortion makes the T* higher for NLNO/NGO. A recent work on the NdNiO₃ films grown on NGO, LSAT and LAO substrates, which possess orthorhombic, cubic and rhombohedral symmetries respectively shows that the transport properties changes drastically [5] and depend on the tilt pattern of oxygen octahedra at film/substrate interface. It has been suggested by imposing the specific tilts of oxygen octahedra using NGO (111) substrate, the MIT can be pushed to 335 K [5].Thus the unconventional direction of substrate used for growing nickelates film changes the transition temperature dramatically.

The observed MCs in NLNO films though very small have distinct behavior that has not been reported before. The MC shows contribution from the weak localization and also a contribution arises from the formation of local moments at energies close to the Fermi level at the approach of MIT. The two contributions add-up and give the non-trivial temperature dependence of the MC. Both the components are enhanced as $\sigma_0$ decreases and the MIT is approached. Appearance of local moments close to the critical region is an important hallmark of Anderson Transition [39]. The MC result reinforces the inference that the substitution leads to a regime of Anderson type transition.

Our result shows that transport properties have dominant contributions from strain, as well as crystal lattice symmetry. The strain is responsible for tuning the transition from compressive (NLNO/LAO) to tensile strain (NLNO/STO or NLNO/NGO) but the substrate symmetry and orientation plays an important role for the case where transport is investigated on both films experiencing tensile strain (NLNO/STO and NLNO/NGO in this case).This, however masks any systematic dependence on disorder. Nevertheless, there exists a role of disorder in the films. Investigation of quenched disorder in $La_{0.5}Eu_{0.5}NiO_3$ shows [11] that strain along with quenched disorder can significantly alter the carrier dynamics in the nickelates. Our investigation reinforces that statement and shows that all these factors can change the nature of the MIT significantly and can make the approach to the transition a continuous one. However a further detailed crystallographic analysis is necessary to understand the effect of strain and lattice symmetry on the $NiO_6$ octahedra in NLNO films. Also, controlled introduction of disorder retaining other factors unchanged would clarify the role of disorder in MIT of such systems.

## 5. Conclusion

In this paper we show that the first order MIT (like a Mott transition) in $NdNiO_3$ change its nature when Nd is substituted by La close to the critical region of a composition driven MI transition as in an Anderson type transition. The distinct MI transition temperature $T_{MI}$ changes to a crossover temperature ($T^*$) that marks change from a PTC of resistivity at $T^* >$ T to a NTC of resistivity at lower T. The transition can further tuned by a combination of factors like strain, substrate symmetry, and quenched disorder. Probes like structural study and introduction of controlled disorder would help further to understand the detailed underlying mechanism of such phenomenon.

**Acknowldgements:**

The authors thank Science and Engineering Research Board (SERB), India for a sponsored project (EMR/2016/002855/PHY). AKR acknowledges additional financial support from J.C. Bose Fellowship of SERB (SR/S2/JCB-17/2006).

**References:**

1. Torrance J B, Lacorre P, Nazzal A I, Ansaldo E J, and Niedermayer Ch 1992 *Phys. Rev.* B 45 8209
2. Catalan G 2008 *Phase Trans.* 81 729
3. Scherwitzl R, Gariglio S, Gabay M, Zubko P, Gibert M, and Triscone J M 2011 *Phys. Rev. Lett.* 106 246403
4. Scherwitzl R, Zubko P, Lichtensteiger C, and Triscone J M 2009 *Appl. Phys. Lett.* 95 222114
5. Catalano S, Gibert M, Fowlie J, Íñiguez J, Triscone J M, and Kreise J 2018 *Rep. Prog. Phys.* 81 046501
6. Heo S, Oh C, Eom M J, Kim J S, Ryu J, Son J, and Jang H M 2016 *Sci. Rep.* 6 22228
7. Kumar Y, Choudhary R J, and Kumar R 2012 *J. Appl. Phys.* 112 073718
8. Catalan G, Bowman R M, and Gregg J M 2000 *Phys. Rev.* B 62 7892
9. Mikheev E, Hauser A J, Himmetoglu B, Moreno N E, Janotti A, Van de Walle C G, and Stemmer S 2015 *Sci. Adv.* 1 e1500797
10. Zhang J Y, Kim H, Mikheev E, Hauser A J and Stemmer S 2016 *Sci. Rep.* 6 23652

11. Das S, Phanindra V E, Philip S S, and Rana D S 2017 *Phys. Rev.* B 96 144411

12. Lian X K, Chen F, Tan X L, Chen P F, Wang L F, Gao G Y, Jin S W, and Wu W B 2013 *Appl. Phys. Lett.* 103 172110

13. Stewart M K, Liu J, Kareev M, Chakhalian J, and Basov D N 2011 *Phys. Rev. Lett.* 107 176401

14. Bisht R S, Samanta S, and Raychaudhuri A K 2017 *Phys. Rev.* B 95 115147

15. García-Muñoz J L, Aranda M A G, Alonso J A and Martínez-Lope M J 2009 *Phys. Rev.* B 79 134432

16. Blasco J and Garcia J 1994 *J. Phys.: Condens. Matter* 6 10759-10772

17. Disa A S, Kumah D P, Ngai J H, Specht E D, Arena D A,Walker F J, and Ahn C H 2013 *APL Mater.*1 032110

18. Möbius A 2017 *Critical Reviews in Solid State and Materials Sciences* 0 1-55

19. Raychaudhuri A K 2006 *Adv. Phys.*44 21

20. Rajeev K P and Raychaudhuri A K 1992 *Phys. Rev.* B 46 1309

21. Raychaudhuri A K, Rajeev K P, Srikanth H and Mahendiran R 1994 *Physica* B 197 124-132

22. Mott N F1990 (Taylor & Francis, London)

23. Imada M, Fujimori A, and Tokura Y 1998 *Rev. Mod. Phys.* 70 1039

24. Son J, Moetakef P, LeBeau J M, Ouellette D, Balents L, Allen J S, and Stemmer S 2010 *Appl. Phys. Lett.* 96 062114

25. Rajeev K P, Shivashankar G V and Raychaudhuri A K 1991 *Solid State Commun.*79 591

26. Moon E J , Gray B A , Kareev M , Liu J , Altendorf S G , Strigari F, Tjeng L H, Freeland J W and Chakhalian J 2011 *New J  Phys* 13 073037

27. Herranz G, Martinez B, Fontcuberta J, Sanchez F, Ferrater C, Garcia-Cuenca M V, and Varela M 2003 *Phys. Rev.* B 67 174423

28. Shi Y G, Guo Y F, Yu S, Arai M, Belik A A, Sato A, Yamaura K, Takayama-Muromachi E, Tian H F, Yang H X,  Li J Q, Varga T, Mitchell J F, and Okamoto S 2009 *Phys. Rev.* B 80 161104

29. Raychaudhuri A K 1991 *Phys. Rev.* B 448572

30. Efros A L, Pollak M 1985 Electron-Electron interaction in disordered systems

31. Patel N D, Mukherjee A, Kaushal N, Moreo A, and Dagotto E 2017 *Phys. Rev. Lett.* 119 086601

32. Oliveira W S, Aguiar M C O, and Dobrosavljevic V 2014 *Phys. Rev.* B 89 165138

33. Wang L, Ju S, You L, Qi Y, Guo Y, Ren P, Zhou Y, and Wang J 2015 *Sci. Rep.* 5, 18707

34. Kumar Y, Choudhary R J, Sharma S K, Knobel M, and Kumar R 2012 *Appl. Phys. Lett.* 101 132101

35. Scherwitzl R, Metal-insulator transitions in nickelate heterostructures, PhD thesis, University of Geneva

36. Shannon R D 1976 *ActaCryst.* A **32**(5) 751

37. Shafarman W N, Koon D W, and Castner T G 1989 *Phys. Rev.* B 40 1216

38. Mott N F and Kaveh M 1983 *Philosophical Magazine B* 47 577-603

39. Lee P A and Ramakrishnan T V 1985 *Rev. Mod. Phys.* 57 287

40. Gati E, Tutsch U, Naji A, Garst M, Köhler S, Schubert H , Sasaki T  and  Lang M 2018 *Crystals*  8 38

41. Wolf E L 1989 (Oxford University Press)

42. Kawabata A 1980 *J. Phys. Soc. Jpn.* 49 628

43. Frydman A and Ovadyahu Z 1995 *Solid State Commun*. 94 745

44. Vaknin A, Frydman A, Ovadyahu Z, and M. Pollak 1996 *Phys. Rev.* B 54 13604